\documentclass{aastex61}

\newcommand\aastex{AAS\TeX}

\submitjournal{ApJ}

\shorttitle{\aastex\ RR Lyrae in the Galactic Centre}
\shortauthors{Contreras Ramos et al.}

\begin{document}

\title{The VVV Survey RR Lyrae Population in the Galactic Centre Region\footnote{Based on observations taken with ESO telescopes at Paranal Observatory under programme IDs 179.B-2002}}

\correspondingauthor{Rodrigo Contreras Ramos}
\email{rcontrer@astro.puc.cl}
\affil{Instituto Milenio de Astrof\'isica, Santiago, Chile.}
\affil{Pontificia Universidad Cat\'olica de Chile, Instituto de Astrofisica, Av. Vicuna Mackenna 4860, Santiago, Chile.}

\author{Rodrigo Contreras Ramos}
\affil{Instituto Milenio de Astrof\'isica, Santiago, Chile.}
\affil{Pontificia Universidad Cat\'olica de Chile, Instituto de Astrofisica, Av. Vicuna Mackenna 4860, Santiago, Chile.}

\author{Dante Minniti}
\affil{Instituto Milenio de Astrof\'isica, Santiago, Chile.}
\affil{Departamento de Ciencias F\'isicas, Facultad de Ciencias Exactas, Universidad Andr\'es Bello, Av. Fern\'andez Concha 700, Las Condes, Santiago, Chile.}
\affil{Vatican Observatory, V00120 Vatican City State, Italy.}

\author{Felipe Gran}
\affil{Instituto Milenio de Astrof\'isica, Santiago, Chile.}
\affil{Pontificia Universidad Cat\'olica de Chile, Instituto de Astrofisica, Av. Vicuna Mackenna 4860, Santiago, Chile.}

\author{Manuela Zoccali}
\affil{Instituto Milenio de Astrof\'isica, Santiago, Chile.}
\affil{Pontificia Universidad Cat\'olica de Chile, Instituto de Astrofisica, Av. Vicuna Mackenna 4860, Santiago, Chile.}

\author{Javier Alonso-Garc\'ia}
\affil{Unidad de Astronom\'ia, Facultad Cs. B\'asicas, Universidad de Antofagasta, Avda. U. de Antofagasta 02800, Antofagasta, Chile}
\affil{Instituto Milenio de Astrof\'isica, Santiago, Chile.}

\author{Pablo Huijse}
\affil{Instituto Milenio de Astrof\'isica, Santiago, Chile.}
\affil{Instituto de Inform\'atica, Universidad Austral de Chile, General Lagos 2086, Valdivia, Chile}

\author{Mar\'ia Gabriela Navarro}
\affil{Instituto Milenio de Astrof\'isica, Santiago, Chile.}
\affil{Departamento de Ciencias F\'isicas, Facultad de Ciencias Exactas, Universidad Andr\'es Bello, Av. Fern\'andez Concha 700, Las Condes, Santiago, Chile.}

\author{\'Alvaro Rojas-Arriagada}
\affil{Instituto Milenio de Astrof\'isica, Santiago, Chile.}
\affil{Pontificia Universidad Cat\'olica de Chile, Instituto de Astrofisica, Av. Vicuna Mackenna 4860, Santiago, Chile.}

\author{Elena Valenti}
\affil{European Southern Observatory, Karl-Schwarszchild-Str. 2, D-85748 Garching bei Muenchen, Germany.}

\begin{abstract}
Deep near-IR images from the VISTA Variables in the V\'ia L\'actea (VVV) Survey were used to search for RR Lyrae stars within 100 arcmin from the Galactic Centre. A large sample of 960 RR Lyrae of type ab (RRab) stars were discovered. A catalog is presented featuring the positions, magnitudes, colors, periods, and amplitudes for the new sample, in addition to estimated reddenings, distances, and metallicities, and measured individual relative proper motions. We use the reddening-corrected Wesenheit magnitudes, defined as $W_{K_s}=K_s-0.428 \times (J-K_s)$, in order to isolate bona-fide RRL belonging to the Galaxy Centre, finding that 30 RRab are foreground/background objects. We measure a range of extinctions from $A_{K_s}=0.19$ to $1.75$ mag for the RRab in this region, finding that large extinction is the main cause of the sample incompleteness. The mean period is $P=0.5446 \pm 0.0025$ days, yielding  a mean metallicity of $[Fe/H] = -1.30 \pm 0.01$ ($\sigma=0.33$) dex for the RRab sample in the Galactic Centre region. The median distance for the sample is $D=8.05 \pm 0.02$ kpc. We measure the RRab surface density using the less reddened region sampled here, finding a density of $1000$ RRab/sq deg at a projected Galactocentric distance $R_G=1.6$ deg. Under simple assumptions, this implies a large total mass ($M>10^9 M_\odot$) for the old and metal-poor population contained inside $R_G$. We also measure accurate relative proper motions, from which we derive tangential velocity dispersions of  $\sigma V_l = 125.0$ and $\sigma V_b = 124.1$ km/s along the Galactic longitude and latitude coordinates, respectively. 
The fact that these quantities are similar indicate that the bulk rotation of the RRab population is negligible, and implies that this population is supported by velocity dispersion. In summary, there are two main conclusions of this study. First, the population as a whole is no different from the outer bulge RRab, predominantly a metal-poor component that is shifted respect the Oosterhoff type I population defined by the globular clusters in the halo.
Second, the RRab sample, as representative of the old and metal-poor stellar population in the region, have high velocity dispersions and zero rotation, suggesting a formation via dissipational collapse.
\end{abstract}


\section{Introduction}
\label{s_intro}
The Milky Way (MW) bulge exhibits a complex structure with composite stellar populations \citep{minniti96}.  Recent studies have found a composite Galactic bulge with at least two populations (metal-rich $[Fe/H] \sim +0.3$ and metal-poor $[Fe/H] \sim -0.4$) having different kinematics and spatial distributions \citep[e.g.][and references therein]{zoccali17,rojasarriagada17}. The different structures coexisting in the inner MW may be due to different formation mechanisms.
But when we want to select the first born “primordial” populations, it is logical to concentrate on the most metal-poor stars. The RR Lyrae (RRL) pulsating variable stars are well known tracers of old and metal-poor populations, often present in globular clusters, in the Galactic halo, and bulge \citep[e.g.][]{smith04,catelan04}. They are excellent distance and reddening indicators \citep{carney95,walker89,longmore90,alcock98,bono01,catelan04,sollima06}, and are found in large numbers in the Galactic bulge \citep{soszynski11,dekany13,pietrukowicz15}. Even though the dominant population of the bulge is metal-rich \citep[e.g.][and references therein]{zoccali17}, its RRL stars are metal-poor, with mean $[Fe/H] \approx -1$ dex \citep{smith84,butler76,rodgers77,walker91}. They also seem to have primordial He abundance, similar to old Galactic globular clusters \citep{marconi18}. As expected, the bulge RRL stars also have hot kinematics, with low rotation and large velocity dispersion \citep[e.g.][]{gratton87,minniti96,kunder16}.
These RRL are therefore ideal probes to study the oldest stars in the inner regions of our Galaxy, close to the Galactic Centre, where large differential extinction and crowding complicates the discrimination of different populations. Indeed, previous near-infrared (NIR) searches managed to find only a dozen RRab in the Galactic Centre region \citep{minniti16,dong17}.

The VISTA Variables in Via L\'actea (VVV) is a deep NIR survey carried out to identify variable stars in some of the most crowded and reddened regions of the Galaxy  \citep[][Hempel et al. submitted]{minniti10,saito12}.  The survey includes the region of the Galactic Centre, enabling the discovery of RRL variables in this region \citep{minniti16}, which are the focus of this study, where we present a new catalog of about a thousand RRL type ab (RRab) stars. This is a vast improvement in the number of known RRL upon previous studies \citep{minniti16,dong17}, and allows the study and characterization of the RRab population in the central most region of our Galaxy.

This paper is organized as follows. Section~\ref{s_selection} describes the selection of the RRL stars, while Section~\ref{s_cmd} presents color-magnitude diagrams (CMDs) for our sample. Section~\ref{s_reddening} discusses the reddening and extinction estimates, whereas in Section~\ref{s_metallicity} we discuss the period distribution and inferred metallicity estimates. The computed distances are provided in Section~\ref{s_distance}, which likewise includes a discussion concerning the sample's spatial distribution, and in Section ~\ref{s_density} we discuss the total density of the old and metal-poor population. Section~\ref{s_pms} presents the proper motions (PMs) and the derived kinematics for the sample. Section~\ref{s_populations} presents a comparison with other RRL populations. Finally, the conclusions are summarized in Section~\ref{s_conclusion}.
\\

\section{The VVV Survey RR Lyrae Selection} 
\label{s_selection}
The VVV survey has been acquiring data since early 2010 using the VISTA InfraRed Camera (VIRCAM) mounted in the VISTA telescope in Cerro Paranal Observatory, Chile. VIRCAM is a 16-detector array of $2048 \times 2048$ pixels each, with a pixel scale of 0.34 arcsec/pix \citep{dalton06,emerson10}. The VVV observation schedule includes single-epoch photometry in $ZYJH$ bands and a variability campaign in $K_{s}$ filter \citep{minniti10}. At each sky pointing, six dithered images called pawprints are acquired and then combined in a single ``tile'', a mosaic where the large gaps between the pawprint detectors are filled. The entire VVV observations comprise 348 tiles, 194 tiles in the bulge and 152 in the disk area \citep{saito12} each covering a Field of View (FoV) of $\sim1.6$ sq degrees. Public aperture photometric catalogs for all the regions covered by the VVV survey are provided by the Cambridge Astronomical Survey Unit (CASU) using the VISTA Data Flow System Pipeline \citep{emerson04,irwin04}, however, in this work we focus on the innermost region of the Galaxy, where the crowding is so severe that PSF modelling is mandatory. Accordingly, the photometric reduction of each detector was carried out using the DAOPHOT II/ALLSTAR package \citep{stetson87} and the CASU catalogs were used to calibrate our photometry into the VISTA system by means of a simple magnitude shift using several thousands stars in common \citep{contrerasramos17}. We applied this procedure separately on each detector of the VVV tiles surrounding the inner $100$ arcmin across the Galactic Centre, comprising tiles b304, b305, b318, b319, b320, b332, b333, b334, b346, b347 and b348. The reduced data include $\sim100$ epochs spanning six years (2010-2015) of observations.

Searching for relatively faint and low amplitude variable stars in the centre of the MW, as is the case of RRL stars, is hampered by extremely high extinction and crowding. However, the dust in this central region of our Galaxy is far from being homogeneously distributed \citep[see for example][and references therein]{gonzalez12,alonsogarcia17}, meaning that RRL stars residing in the very inner bulge show quite different apparent magnitudes (but very similar reddening-corrected ones). Thus, those RRL less affected by extinction are well detectable using VVV data, as revealed in our initial search \citep{minniti16}. 

RRL stars are commonly found into two main types, RRab which pulsate in the fundamental mode and have longer periods and typically higher amplitudes than RR type c (RRc), that are first overtone pulsators. Part of the difficulty for finding RRL in the NIR is that their amplitudes of variation are smaller than in the optical, typically $A(K_s) \sim 1/3 A(V)$ \citep[e.g.][]{angeloni14,navarrete15,navarrete17,braga18}. The identification of RRc stars is even trickier because their light curves are rather sinusoidal, not very different from the morphology found in short-period eclipsing binaries. For these reasons, we focus our search on RRab only, which feature a unique saw-tooth light curve, longer periods and higher amplitudes. Their identification was made using visual inspection and two relatively new automated RRab classifiers, one described in \cite{elorrieta16} and a new algorithm especially developed for this work (see appendix A). Their periodicities were obtained by means of the analysis of variance statistic \citep{schwarzenbergczerny89} in the RRL period range. Representative light curves of our RRab sample can be seen in Figure~2 of \cite{minniti16}.

\begin{figure}[t]
\centering
\includegraphics[width=0.5\hsize]{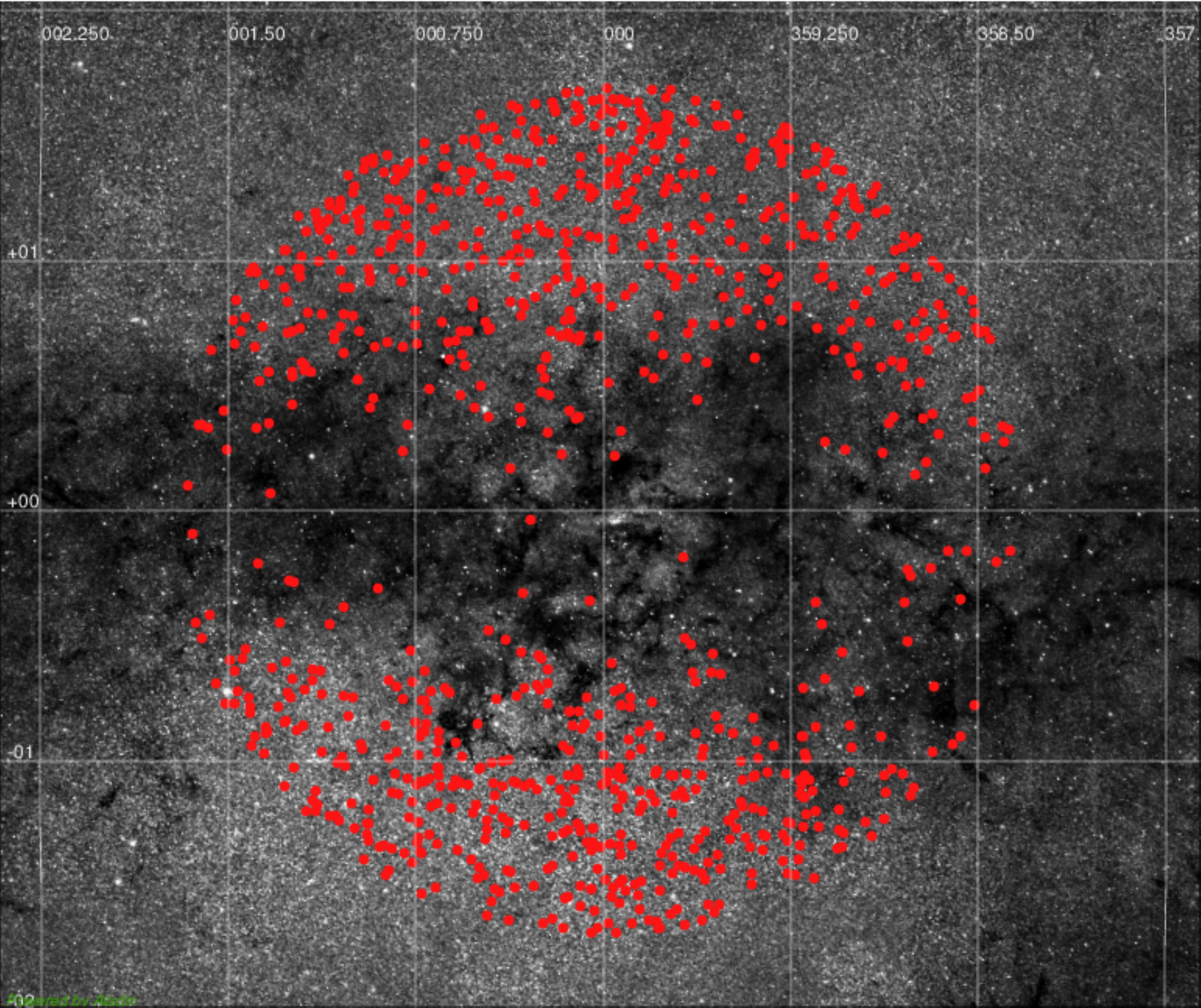}
\caption{Location in Galactic coordinates of the new RRL candidates stars within a 4$\times$4.5 sq deg map surrounding the Galactic Centre. The red dots mark the individual positions of the 960 RRL candidates located within 100 arcmin of the Galactic Centre discovered with the NIR observations of the VVV Survey. At the distance of the Galactic Centre, the scale is 1 degree $\simeq$ 150 pc. North galactic pole is up, positive galactic longitudes are left.}
\label{f_rrl_map2}
\end{figure}

Figure~\ref{f_rrl_map2} shows the map of the RRab candidates found within 100 arcmin from the Galactic Centre. As can be seen, most of the detected RRL stars are projected in regions of the inner bulge where the extinction is not extremely high. In fact, from the figure it is clear the dearth of RRL in the region $|b|\lesssim 1^\circ$ due to the heavy extinction close to the Galactic plane. The final catalog presented here includes 960 objects in total, of which 948 are new discoveries. 
For each RRL, we provide in Table~\ref{t_tab1} galactic coordinates, NIR magnitudes in the five VVV filters and relative proper motions with their corresponding statistical errors. In Table~\ref{t_tab2} we include extinction parameters, distance values, periods, $K_{s}$ amplitudes and metallicities.

\begin{figure}[t]
\centering
\includegraphics[width=0.4\hsize]{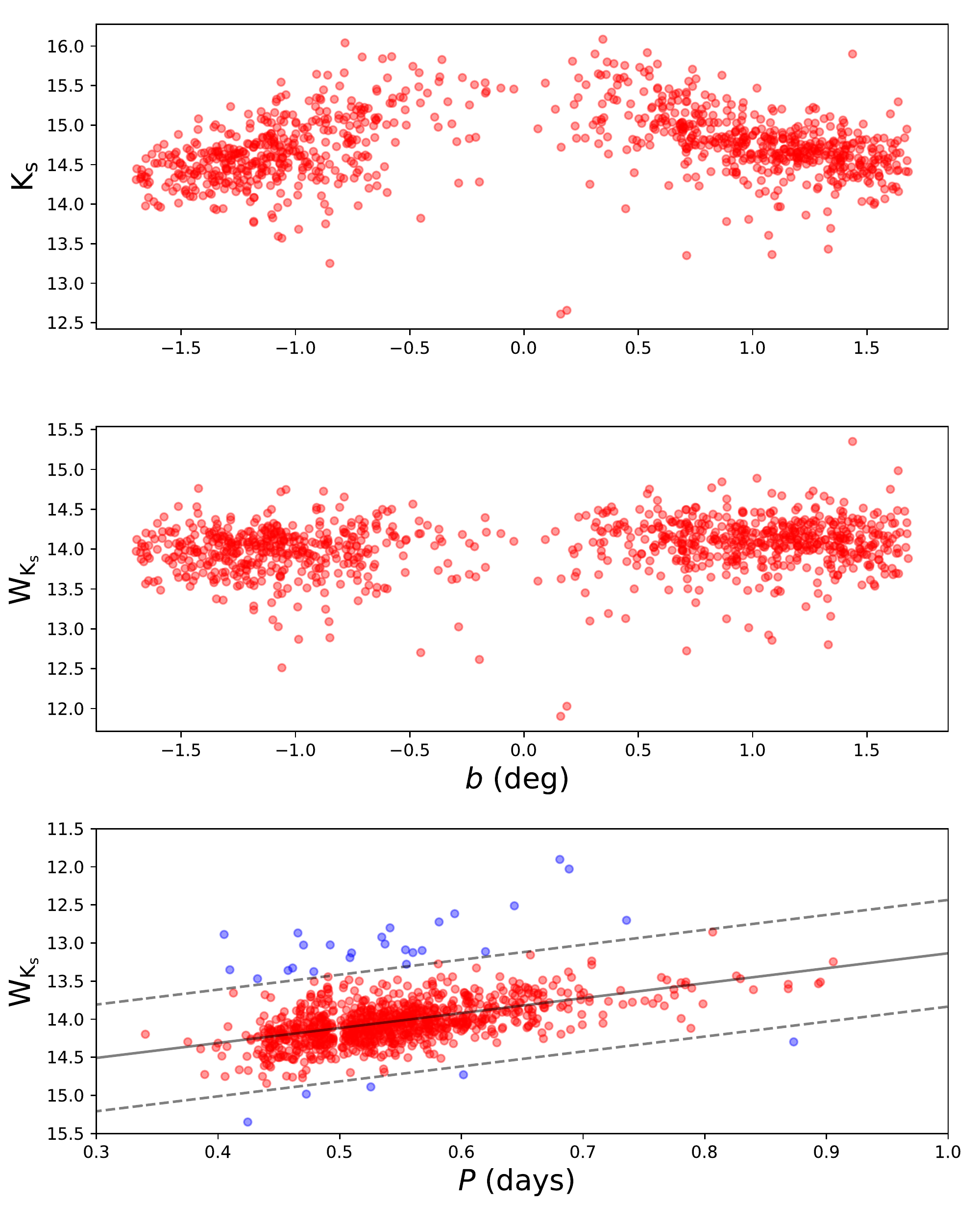}
\caption{Top: $K_{s}$-band magnitude $vs$ Galactic latitude for the 959 RRL candidates located within 100 arcmin of the Galactic Centre. The effect of higher extinction closer to the Galactic plane is clear. Center: Extinction-corrected (Wesenheit) $W_{K_s}$-band magnitude $vs$ Galactic latitude of the sampled variables. Bottom: $W_{K_s}$-band magnitude $vs$ period. This diagram allows us to make a cut in order to separate Galactic Centre RRL (red dots) from foreground/background RRL and other contaminating objects (30 blue dots).}
\label{f_KvsWK}
\end{figure}

The top panel of Figure~\ref{f_KvsWK} shows the $K_{s}$-band magnitude $vs$ Galactic latitude for the 960 RRL candidates, illustrating the effect of extinction. As one approaches the Galactic plane, at latitudes $|b|\lesssim 1^\circ$, the mean $K_s$-band magnitudes get fainter, until they fall beyond our detection limit at $K_{s} \sim 16$ mag. Again, the lack of RRab in the Galactic plane is caused by heavy extinction, and our sample is severely incomplete in the regions where $E(J-K_{s}) \gtrsim 3$ mag. The middle panel of Figure~\ref{f_KvsWK} shows the same figure using the reddening-corrected Wesenheit magnitudes, defined as $W_{K_s}=K_s-0.428 \times (J-K_s)$, and discussed in detail in Section 4. When the extinction is taken into account, these Wesenheit magnitudes depend on the RRab periods as shown in the bottom panel of Figure~\ref{f_KvsWK}.
After applying a $3\sigma$ clipping to the $P-W_{K_s}$ distribution, this diagram allows us to make a cut in order to separate Galactic Centre RRL stars 
from the foreground and background counterparts. The diagram singles out the presence of 5 distant RRab and of 25 RRab located in front of the Galactic bulge. We also note that there is one star that has no $J-$band photometry.
\\

\section{The Color-Magnitude Diagram}
\label{s_cmd}
\begin{figure}[h]
\centering
\includegraphics[width=0.6\hsize]{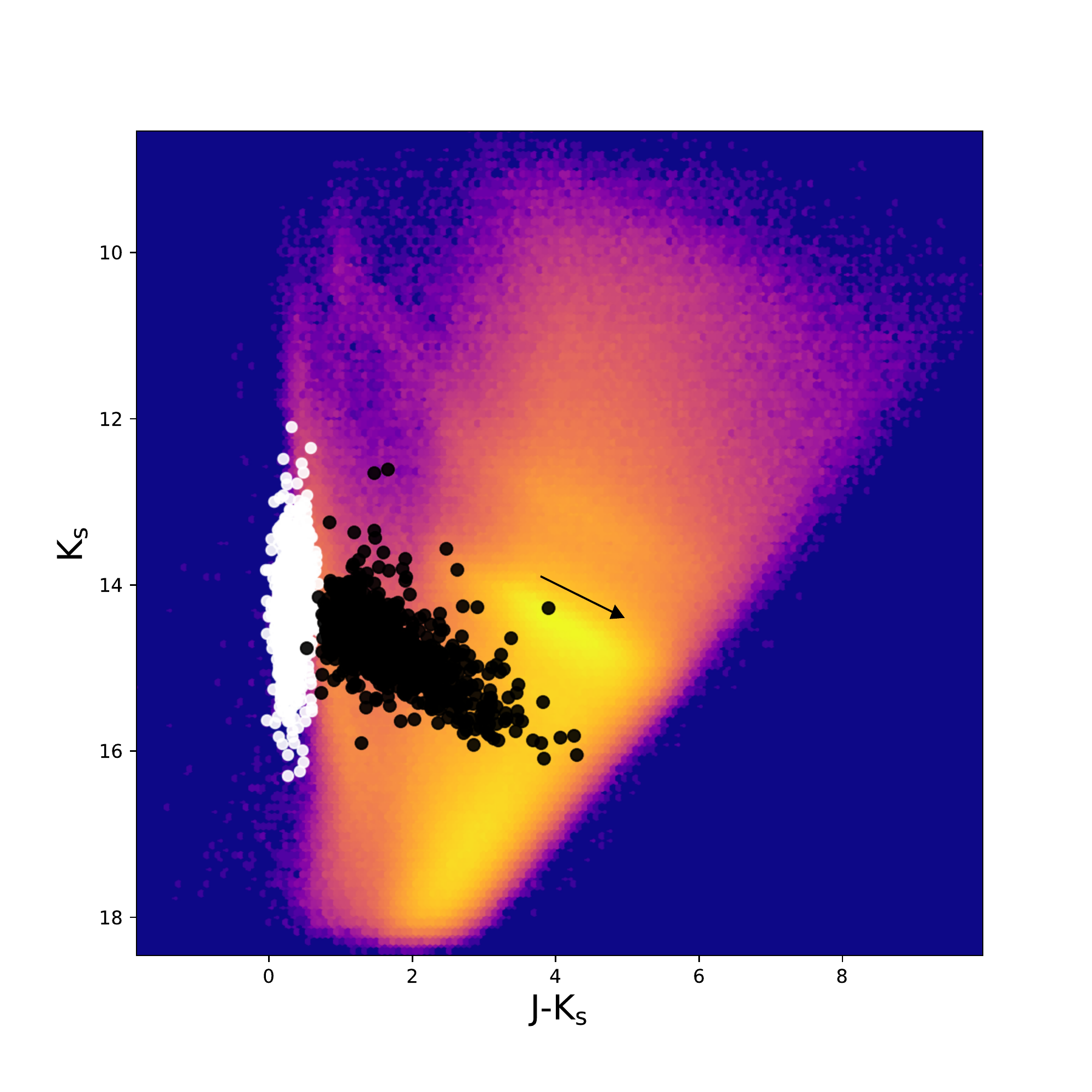}
\caption{VVV $(J-K_{s})$ $vs$ $K_{s}$-band color-magnitude diagram for the Galactic Centre region (tiles b333 and b319) shown as a Hess density diagram \citep{minniti16}. The sample of $\sim1000$ outer bulge RRL from \cite{gran16} is plotted with white dots for comparison. Our 959 RRL candidates located within 100 arcmin of the Galactic Centre are plotted as black dots. The reddening vector with slope $\Delta K_{s}/ \Delta (J-K_{s})= 0.428$ is shown with the arrow \citep{alonsogarcia17}.}
\label{f_cmd}
\end{figure}

The observed $K_{s}$ $vs$ $J-K_{s}$ color-magnitude diagram (CMD) of the VVV tiles b333 and b319, containing the Galactic Centre and part of the area studied here, is shown in Figure~\ref{f_cmd}. This region contains $\sim7$ million point sources, that are plotted as a density diagram. Our sample of 959 RRL candidates (one RRab has no $J$-band photometry) is overlaid in the CMD as black dots. For comparison, the sample of $\sim1000$ outer bulge RRab from \cite{gran16} is also plotted (white dots). The latter is almost unreddened because the outer bulge fields have $E(J-K_{s})<0.2$ mag, and the former is a more differential reddened population that lies along the direction of the reddening vector, as expected. The reddening vector shown in  Figure~\ref{f_cmd} with the black arrow has a slope $\Delta K_{s} / \Delta (J - K_{s}) = 0.428$, following the new results of \cite{alonsogarcia17} obtained using VVV data in the inner region of the Galaxy.

\begin{figure}[h]
\centering
\includegraphics[width=1.0\hsize]{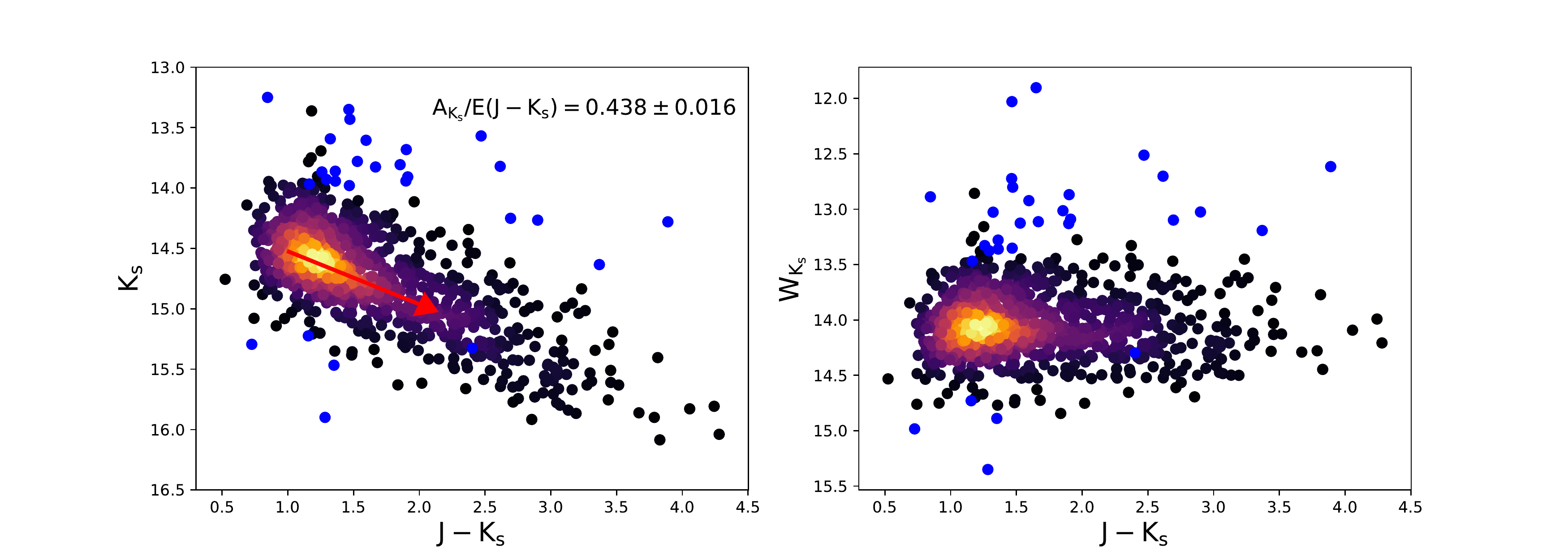}
\caption{Left: VVV $K_{s}$-band $vs$ $(J-K_{s})$ color-magnitude diagram for the 959 RRL candidates located within 100 arcmin of the Galactic Centre. RRL located at the Galactic Centre distance are plotted as a Hess diagram, while blue dots are foreground/background objects. Right: Same as the left panel but using the extinction-corrected (Wesenheit) $W_{K_s}$-band magnitude. This was computed assuming a slope of $\Delta K_{s}/ \Delta (J-K_{s})= 0.438$ from a fit shown by the red arrow in the right panel. This slope agrees within the errors with the slope measured by \cite{alonsogarcia15,alonsogarcia17}.}
\label{f_rrl-redd}
\end{figure}

As they are all at approximately the same distance, we have used the location of the RRab stars in the CMD to compute the total-to-selective extinction ratio at the Galactic Centre, that in presence of differential reddening, should follow the slope $\Delta K_{s}/ \Delta (J-K_{s})$. The left panel of Figure~\ref{f_rrl-redd} shows the CMD in the Galactic Centre region, where we use a Hess density diagram for the RRab sample and blue dots to separate the foreground and background objects from the rest of the population. From a simple least square fitting, we obtained  $A_{K_{s}} / E(J-K_{s}) = 0.438 \pm 0.016$ mag, in very good agreement with the values obtained by \cite{alonsogarcia15,alonsogarcia17}. The effect of the extinction is taken into account in the right panel of Figure~\ref{f_rrl-redd}, that shows the $W_{K_s}$ $vs$ $J-K_{s}$ CMD. 
Note that the mean number of observations (different frames where $K_s$ and PMs are measured) is $N=82$ in the $K_{s}$-band, with a range from $N=52$ to $N=104$. Therefore the mean magnitude errors are negligible for the $K_{s}$-band, because we are using the mean of these multiple epochs, resulting in $\sigma K_{s} = 0.0039\pm 0.0009$ mag. However, fewer epochs (typically 2-4) are available for the other passbands.  Therefore the errors in the $J$-band for example are almost an order of magnitude higher in the mean, $\sigma J = 0.020 \pm 0.019$ mag for the sample stars. Even worse, extinction affects the shorter passbands severely, and the mean errors are much larger in the mean $\sigma Y = 0.04$ mag for the whole sample 
These errors, however, have no major effect in the appearance of the CMDs shown in this work, which are rather dominated by extinction effects.
\\

\section{Reddenings and Extinctions}
\label{s_reddening}
\begin{figure}[b]
\centering
\includegraphics[width=0.50\hsize]{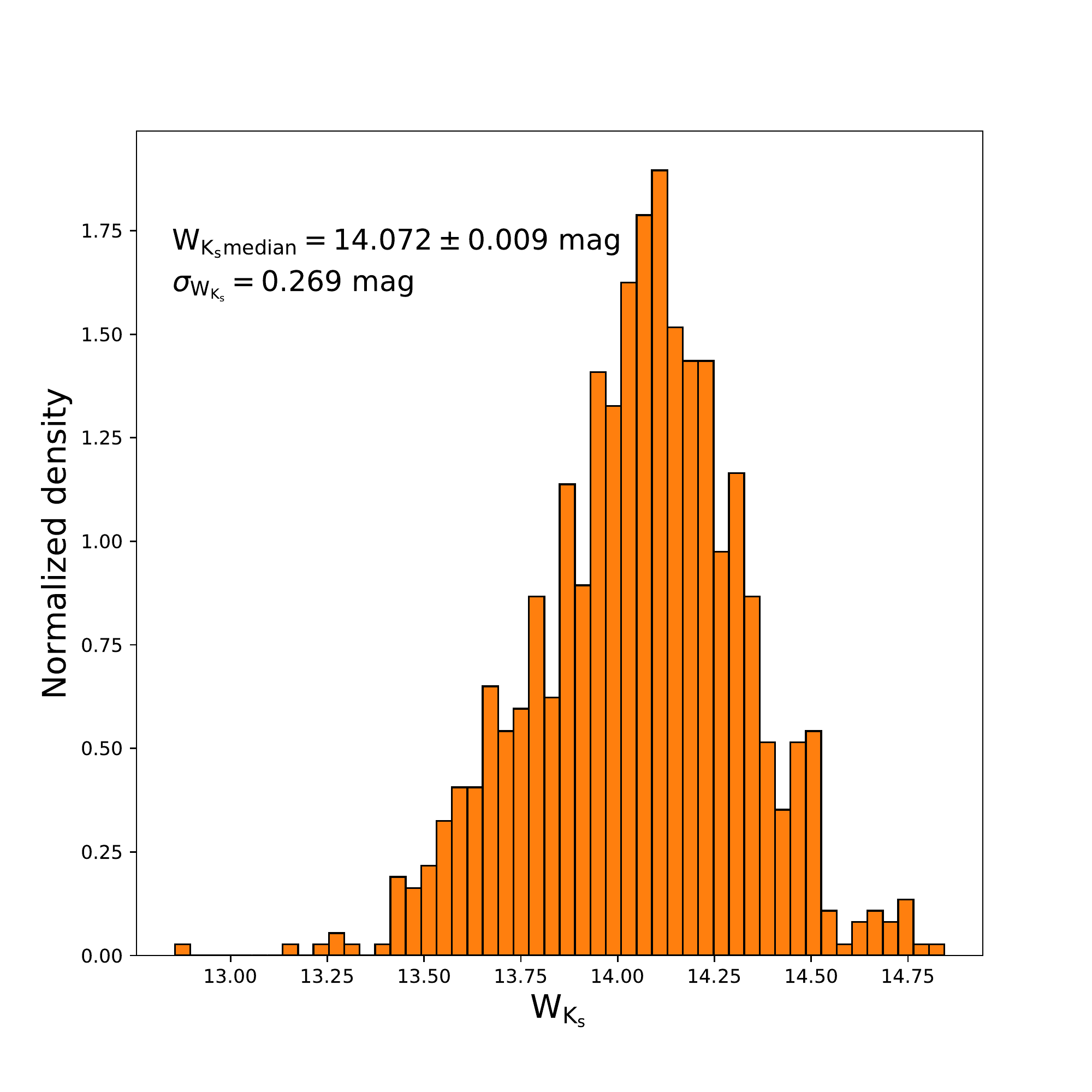}
\caption{The extinction-corrected (Wesenheit) $W_{K_s}$-band magnitude distribution for the 959 RRL candidates located within 100 arcmin of the Galactic Centre, after eliminating foreground/background stars. 
According to the magnitude distribution, the vast majority of the RRL candidates are located at the Galactic Centre region.}
\label{f_histograma_WK}
\end{figure}

The NIR reddening law in the Galactic Centre region has been the subject of a variety of studies \citep[see][and references therein]{matsunaga09,alonsogarcia15,majaess16,nataf16,alonsogarcia17}. What matters for the present study is the slope of this reddening law in the NIR, which we take from the work recently published by \cite{alonsogarcia17}, $A_{K_s} / E(J-K_{s}) = 0.428$. This allows us to define the reddening corrected Wesenheit individual magnitudes $W_{K_s}=K_{s}-0.428 \times (J-K_{s})$ mag for the RRab sample. The final distribution of the $W_{K_s}$ magnitudes, shown in Figure~\ref{f_histograma_WK}, exhibits a strong peak at the expected location of the Galactic Centre. 

\begin{figure}[t]
\centering
\includegraphics[width=0.55\hsize]{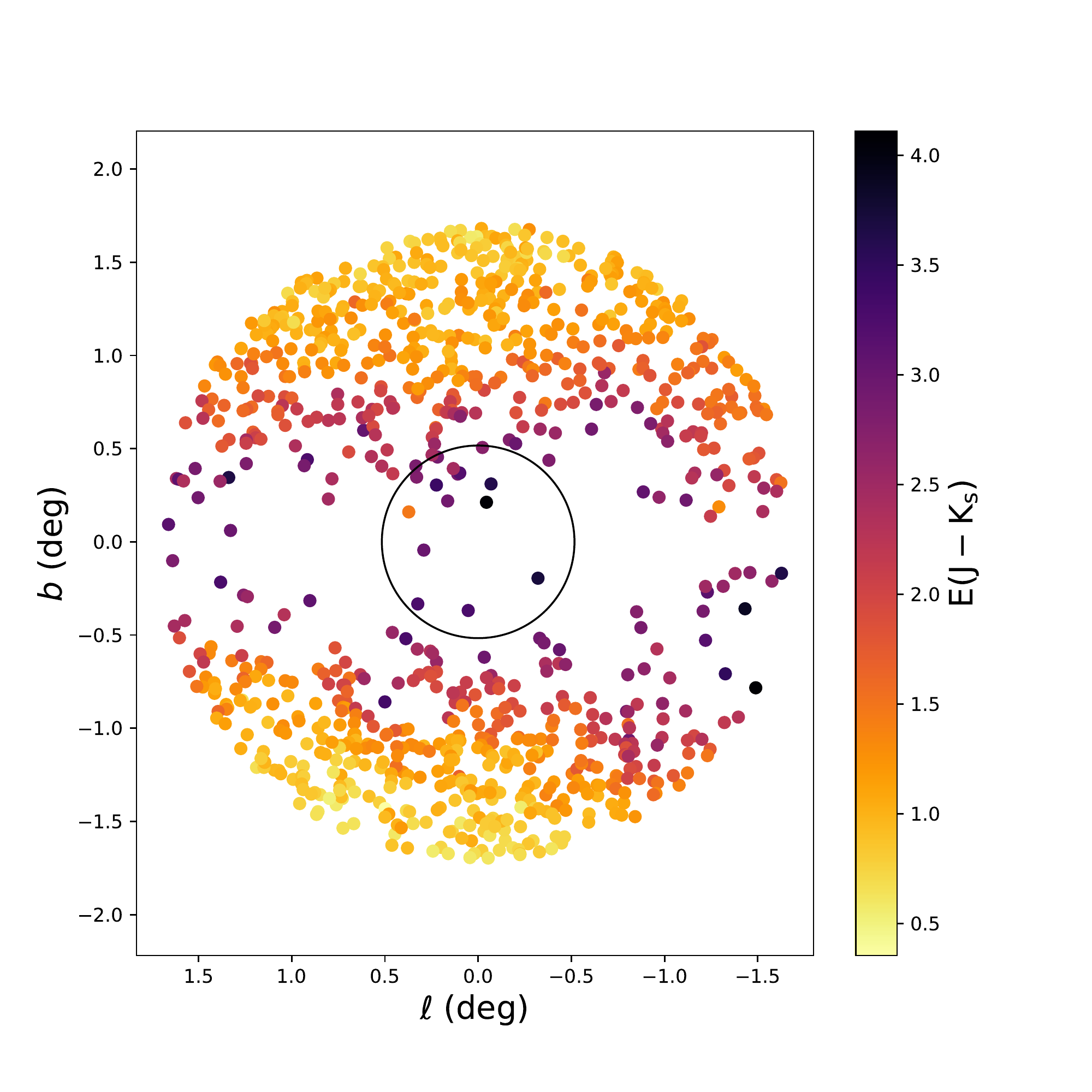}
\caption{Map of the candidate RRL, distinguished by color according to their respective $E(J-K_{s})$. The darkest dots are RRL with colors $J-K_{s}>4$ ($A_{K_s} \sim 1.7$ mag), outlining the most reddened regions. On the other hand, the lightest yellow dots show RRL with colors $J-K_{s}<1.0$, indicating the less reddened regions. The black circle shows the area analyzed by \cite{minniti16}.}
\label{f_extincion_map}
\end{figure}

\begin{figure}[t]
\centering
\includegraphics[width=1.0\hsize]{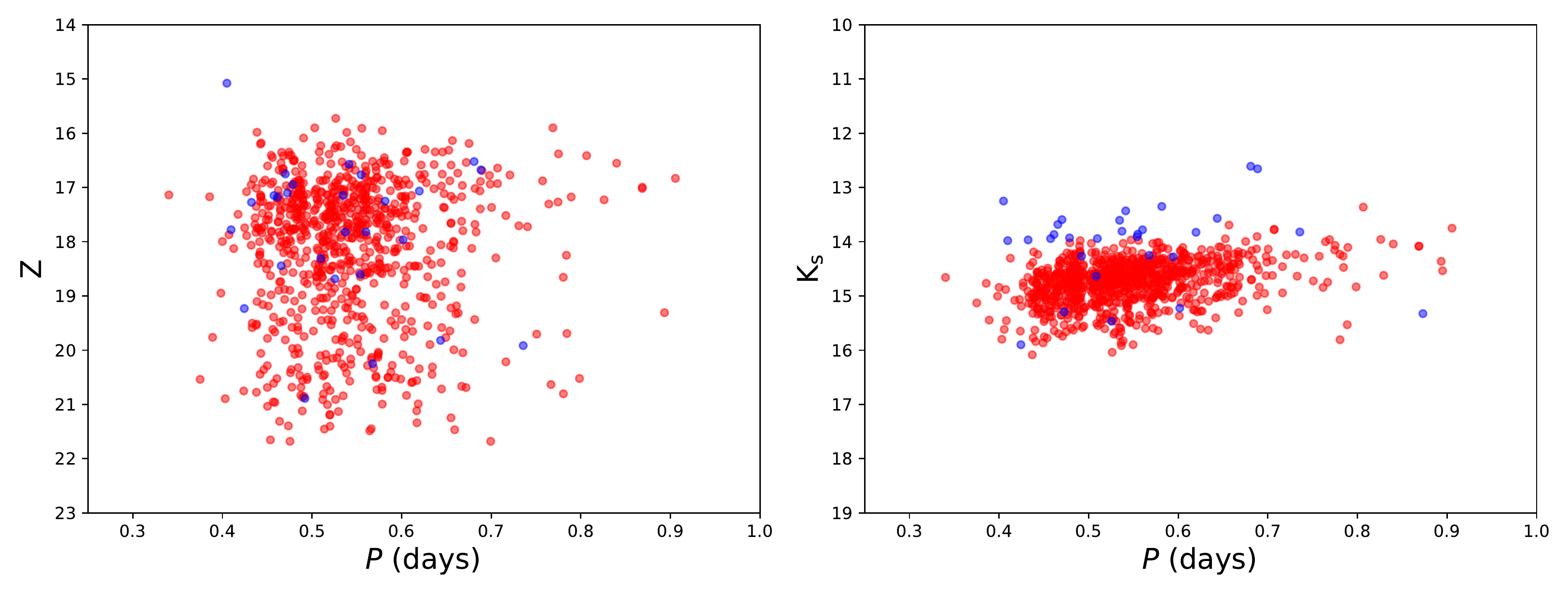}
\caption{Left: $Z$-band magnitude $vs$ period in days for the RRL candidates located within 100 arcmin of the Galactic Centre (red points). 
Right: $K_{s}$-band magnitude $vs$ period in days for our sample. Blue points represent foreground/background objects.} 
\label{f_pl-z-ks}
\end{figure}

Moreover, RRL stars are excellent reddening indicators because in the NIR their temperatures are restricted to a very narrow range in the instability strip. Therefore, it is possible to use their observed colors to estimate reddenings along their line of sight. Based on the data from \cite{navarrete15} we found that the intrinsic color of RRab can be assumed to be $\Delta (J-K_{s})=0.17\pm0.03$ mag. Thus, the individual reddening values listed in Table~\ref{t_tab1} were obtained by the relation $E(J-K_{s})=(J-K_{s})-0.17$ mag.
We inspect the spatial distribution of the bluest and reddest RRab in the sky region surrounding the Galactic Centre. Figure~\ref{f_extincion_map} shows this map for the RRab candidates, color-coded by their respective reddenings. The darkest points represent the most reddened objects with $E(J-K_{s}) > 3.5$ mag (corresponding to $A_{K_s} \gtrsim 1.5$ mag), outlining the most heavily reddened regions close to the Galactic plane. On the other extreme, the lightest yellow dots show the location of the bluest objects with  $E(J-K_{s}) < 1.0$ mag, marking the less reddened regions (corresponding to $A_{K_s} < 0.43$ mag). The bluest/reddest RRab stars of our sample have colors $J-K_{s} \sim 0.61/4.3$ mag, equivalent to $A_{K_s} \sim 0.19/1.75$ mag.
In order to further characterize our sample, we have also examined the magnitude dependence on the RRab periods. Figure~\ref{f_pl-z-ks} (left panel) shows the $Z$-band magnitude $vs$ period in days for our RRL candidates located within 100 arcmin of the Galactic Centre. This figure clearly illustrates the damaging effects of reddening in the passbands at shorter-wavelengths, while this dependence is alleviated for the $K_{s}$-band magnitude (right panel). Even though the scatter due to reddening is still considerable, there is a hint of a period-luminosity relation seen in the $K_{s}$-band diagram. This relation becomes much tighter using the reddening corrected Wesenheit magnitudes $W_{K_s}$, as shown in the bottom panel of Figure~\ref{f_KvsWK}
\\

\section{The Period Distribution and Metallicities}
\label{s_metallicity}
\begin{figure}[t]
\centering
\includegraphics[width=1.0\hsize]{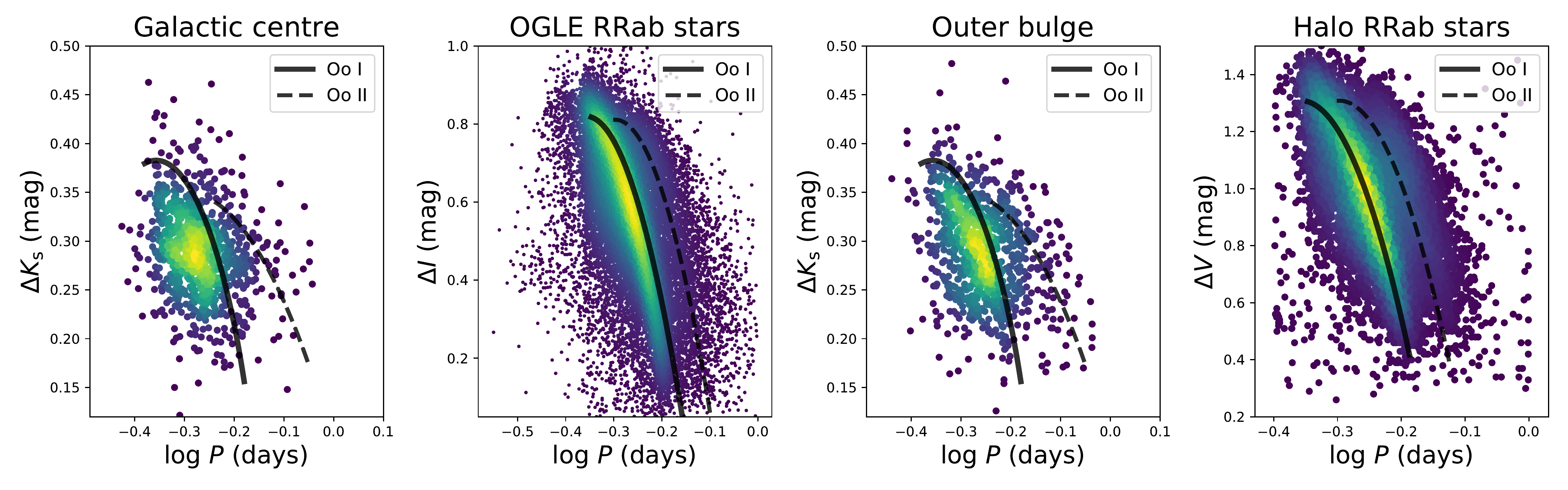}
\caption{Amplitude $vs$ period in days (Bailey diagram) for bulge RRL from different samples. From left to right we show the Bailey diagram for our candidate RRL located within 100 arcmin from the Galactic Centre; all the OGLE RRab stars from \cite{pietrukowicz15}; for the outer bulge sample of 1000 RRab stars from \cite{gran16}; and for the halo RRab from \cite{torrealba15}. The ridge lines of the Oosterhoff I and II populations are indicated in the panels \citep{navarrete15,kunder13,zorotovic10} for comparison. Note that the OGLE and halo samples (2nd and 4th panels) show the amplitudes in the $I$-band and $V$-band respectively, while the present sample and outer bulge sample from the VVV survey show the amplitudes in the $K_s$-band. This figure clearly shows that the Galactic Centre RRL are more extreme than a typical OoI population.}
\label{f_bailey}
\end{figure}

Figure~\ref{f_bailey} shows the $K_{s}$ amplitude $vs$ period in days (a.k.a. Bailey diagram) for the candidate RRab located within 100 arcmin from the Galactic Centre, compared with 1) the $I$-band Bailey diagram sampling $\sim$ 27,000 OGLE RRab stars located in the middle bulge from \cite{pietrukowicz15}, 2) the outer bulge sample of $\sim$ 1000 RRab stars from \cite{gran16} in the $K_{s}$-band, 3) the outer halo sample of $\sim 10,000$ RRab stars published by \cite{torrealba15} in $V$-band. As can be seen in the figure, the three bulge distributions are similar, and no major differences are found. The ridge lines of the Oosterhoff I and II (OoI, OoII) population from \cite{navarrete15}, \cite{kunder13} and \cite{zorotovic10} for the $K_{s}VI$ bands respectively are indicated, showing that the bulk of Galactic bulge RRab stars, independent of their location, do not follow the same locus of the MW globular clusters, and in general are shifted to shorter periods for a given amplitude. It is likely that this period-shift is because bulge RRL stars are in general more metal rich than their counterparts in globular clusters. Similar results were obtained by \cite{kunder09} from their study of bulge RRL stars using MACHO data. On the other hand, outer halo RRab stars nicely match the OoI line, like the majority of relatively metal-rich globular clusters RRab stars.
Quantitatively, the mean periods of the different samples shown in Figure~\ref{f_bailey} are: 
$P_{ab}$ =$0.5446\pm0.0025$ days for our sample of Galactic Centre RRL, 
$P_{ab}$ =$0.5571\pm0.0005$ days for the OGLE bulge sample from \cite{pietrukowicz15}, 
$P_{ab}$ =$0.5634\pm0.0028$ days for the outer bulge RRL from \cite{gran16}, and 
$P_{ab}$ =$0.5751\pm0.0007$ days for the halo RRL sample from \cite{torrealba15}, where the error is the standard error of the mean. These mean periods clearly decrease with decreasing distance to the Galactic Centre, probably indicating that the innermost RRL stars are more metal-rich in the mean.

\begin{figure}[h]
\centering
\includegraphics[width=1.0\hsize]{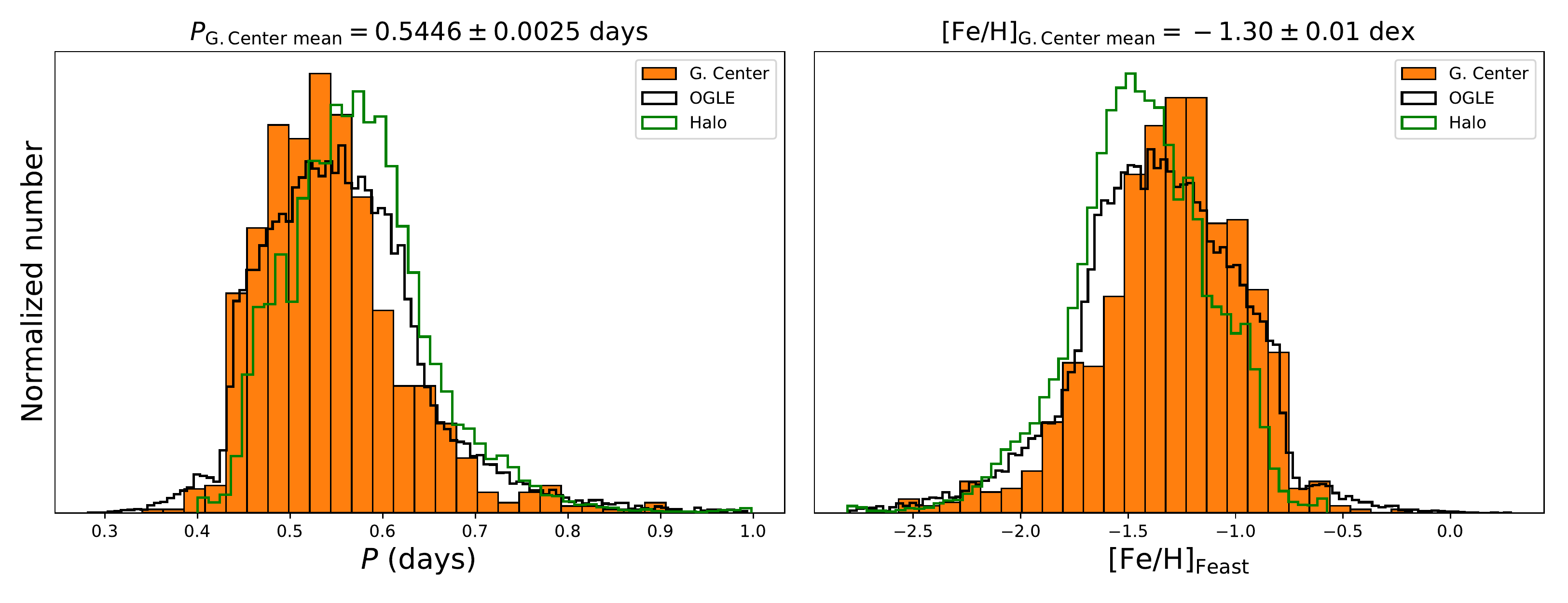}
\caption{Left: Period distribution for our RRab sample compared with OGLE data \citep[black][]{pietrukowicz15} and Outer halo RRab stars \citep[green][]{torrealba15}. There appears to be a mild gradient in period but the differences are small. Right: same as left panel but for the metallicity distribution. The $[Fe/H]$ has been derived using the relation of \cite{feast10}. The mean and the corresponding standard error of the period and metallicity for our RRab sample are indicate in the top of the panels.} 
\label{f_hist_p_met}
\end{figure}

Our Bailey diagram can also be compared with the OGLE Bailey diagram to estimate the completeness of our RRab sample. The NIR light curve amplitudes are lower than in the optical, by more than a factor of 2. In particular, we have very few ($<1$\% of the total sample) low amplitude RRab in our sample with $A(K_s)<0.15$ mag, while the OGLE sample still includes about 10\% of their low amplitude population with $A(I)<0.3$ mag. From this simple comparison we can conclude that our sample is 10\% incomplete, because we are missing  low-amplitude RRab variable stars. The period distribution for our RRL sample is shown in the left panel of Figure~\ref{f_hist_p_met}. 

The RRab periods can be used to estimate individual metallicities for RRab stars, taking for example the RRab stars of \cite{layden94} as calibrators. Different authors have used this relations, which we have also applied for the VVV survey RRab in the plane \citep{minniti17_2}. We use here the relations of \cite{feast10} : $[Fe/H] = -5.62\,\log P_{ab} - 2.81$, and \cite{yang10} : $[Fe/H] = -7.82\,\log P_{ab} - 3.43$. Using these relations we obtain a mean metallicity of $[Fe/H] = -1.30 \pm 0.01$ ($\sigma=0.33$) dex, and $-1.33 \pm 0.02$ ($\sigma=0.46$) dex, respectively. The right panel of Figure~\ref{f_hist_p_met} shows the metallicity distribution obtained using the relation of \cite{feast10}, that is adopted for consistency with our previous work \citep{minniti16, minniti17_2}, and because it gives a smaller dispersion. The individual metallicities estimated using the periods are uncertain \citep{yang10,feast10}, and spectroscopic measurements are highly desirable. While the mean is metal-poor, the derived metallicity distribution is very wide, and allows us to confirm that the sample contains both metal-poor and metal-rich RRab, consistent with the Bailey diagram that shows a mixture of Oosterhoff types I and II populations. In comparison, the periods for the OGLE sample yields an almost identical mean metallicity, $[Fe/H] = -1.35$ dex using the relation of \cite{feast10}. Also, the outer bulge population and outer halo RRab sample give similar metallicities ($[Fe/H] = -1.38$ dex and $[Fe/H] = -1.44$ dex). There appears to be a small gradient in the mean abundances, being RRL in the outer halo more metal-poor in comparison with the Galactic Centre, but the systematics are big and the difference is within the errors. Moreover, we do not observe a significant metallicity gradient in the RRab population within $1.5 > R_G > 0$ kpc.
%
\\

\section{Distances}
\label{s_distance}
\begin{figure}[h]
\centering
\includegraphics[width=0.6\hsize]{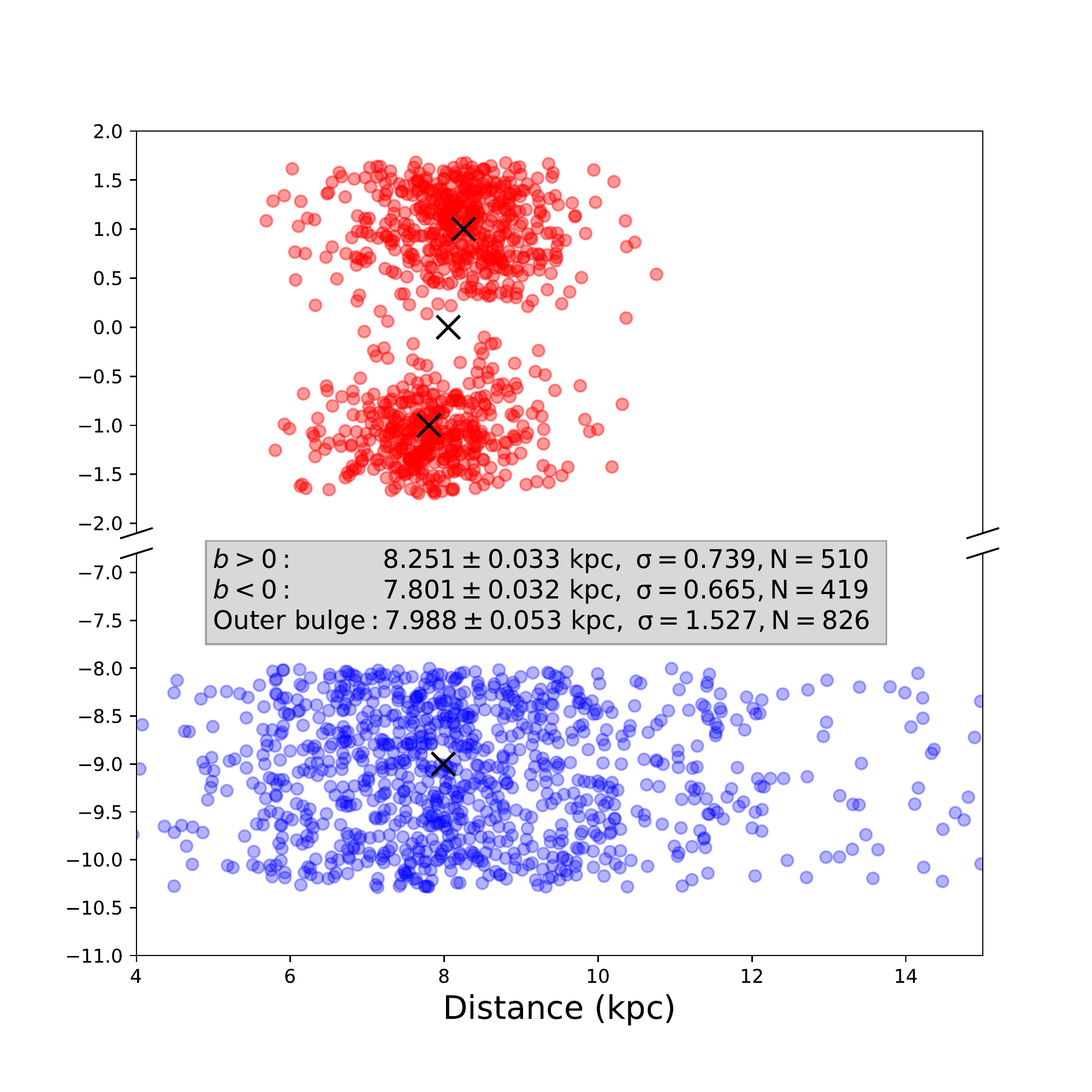}
\caption{Distance distribution for the candidate RRL, separated by Galactic latitude. The distances were computed considering those RRL that most likely belong to the Galactic bulge (stars between 5 and 11 kpc), and the median values of each subsample are indicated (from bottom to top the outer bulge, and the Galactic Centre negative and positive latitude, respectively). The median distances agree within the sigmas of the distributions. This figure shows how concentrated the RRL are at the Galactic Centre.}
\label{f_distances}
\end{figure}

In this section we derive distances for our RRL sample, following the procedures described by \cite{gran16}, and \cite{minniti16}. Briefly, we use the extinction law recently derived by \cite{alonsogarcia17}, the Period-Luminosity (PL) calibration of \cite{muraveva15}: $MK_{s} = -2.53\,\log P_{ab} - 0.95 + 0.07\,\ [Fe/H]$ and we adopted $[Fe/H] = -1.3$ dex as a good aproximation of the individual metallicities (see section \ref{s_metallicity}). Even if the estimated abundance for our RRL stars may be highly uncertain, the PL relation shows a very mild dependence on it. We also note that the impact in the computed distances due to the reduced number of $J-band$ epochs is negligible.
Figure~\ref{f_distances} shows the distance distribution for the candidate RRL, separated by Galactic latitude, compared with the outer bulge RRab from \cite{gran16}. This comparison is appropriate because all these samples have similar systematics, coming from the NIR VVV search for RRL. The median distance and the corresponding errors (standard error) of each subsample is indicated (from bottom to top, the outer bulge and the Galactic Centre negative and positive latitude, respectively). The figure shows how concentrated are the Galactic Centre RRL: their dispersion is much smaller than the $\sigma$ of the RRab located in the outer bulge. The median distance for the complete inner sample of 929 RRab stars is $D=8.052 \pm 0.024$ kpc ($\sigma=0.73$) and the median distance for the RRab sample of \cite{gran16}, 
adopting $A_{k_s}/E(J-K_s)=0.689$ from \cite{cardelli89} and a mean metallicity $[Fe/H] = -1.382$, is $D=8.067 \pm 0.06$ kpc ($\sigma=1.85$). These distances agree within the errors of the distributions. 
\\

\section{Densities}
\label{s_density}
In this section we derive the number density for our RRL sample. This is important because it can give us an idea of how relevant is the old and metal poor population in the Galactic Centre region, where other populations (of young and intermediate age) are also present. In order to have a well known example for comparison, we adopt the numbers corresponding to the giant globular cluster (or dwarf galaxy nucleus) $\omega$ Centauri, for which \cite{navarrete15} presents a complete catalog of RRL stars using similar NIR data obtained with the VIRCAM at the VISTA telescope. Although the mean metallicity of $\omega$ Centauri and bulge RRL stars are comparable, the overall color distribution of the HB stars in the central region of the Galaxy is not known. Therefore we cannot guarantee that the bulge HB stars resemble $\omega$ Centauri HB stars, because of the difficulty of identifying HB stars in the region. Thus, we stress that the comparison with $\omega$ Centauri may be uncertain. According to  \cite{navarrete15}, there are $N=88$ RRab in this cluster \citep{navarrete15}, that has a mass of $M=4\times 10^6 ~M_\odot$ \citep{dsouza13}. That is, there are $\sim45,500 ~M_\odot$ per RRab in this globular cluster.

\begin{figure}[h]
\centering
\includegraphics[width=0.60\hsize]{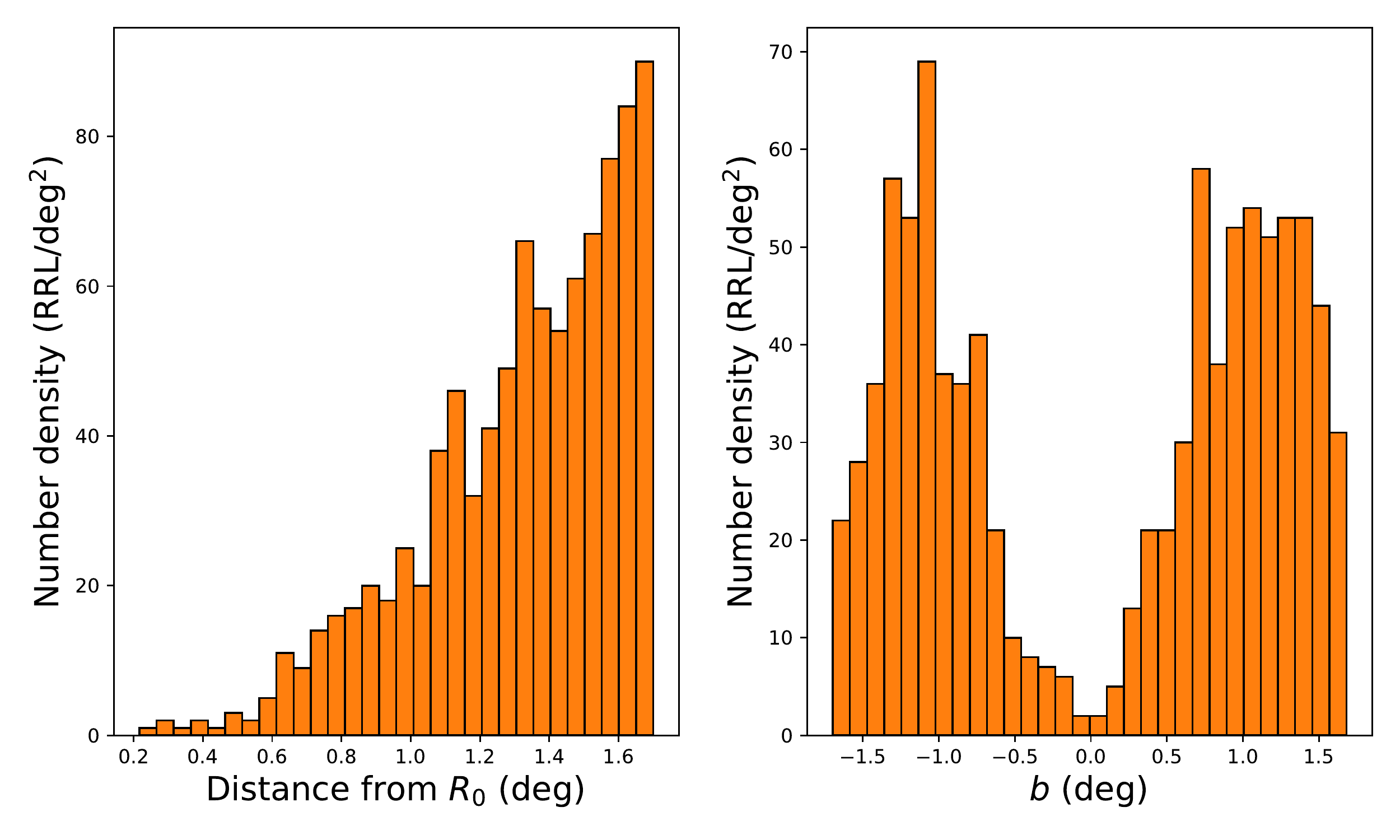}
\caption{Left: Raw candidate RRL number density $vs$ distance from the Galactic Centre in degrees. We find only about 270 RRL within 1 degree of the Galactic Centre, most of the sample stars are located between 60 and 100 arcmin from the Galactic Centre. Right: RRL number density across a strip with $-0.3<l<0.3$ deg perpendicular to the Galactic plane. The less reddened regions away from the plane are more complete, reaching an observed total density of $\sim 1000$ RRL/sq deg at $b=1.6$ deg. However, the number density of RRL drops drastically with decreasing Galactic latitude, where the sample becomes very incomplete due to extinction.}
\label{f_densidad_rrl}
\end{figure}

Figure~\ref{f_densidad_rrl} shows the RRab number counts as function of projected Galactocentric distance in degrees (left panel), and across a narrow strip $-0.3<l<0.3$ deg perpendicular to the Galactic plane (right panel). Based on the less reddened region sampled here we estimate a total density of $\sim 1000$ RRL/sq deg at a Galactocentric distance of $R_G = 1.6$ deg. Using the number ratio of RRab to total mass measured for $\omega$ Centauri, this implies a total density of old and metal-poor stars of $\sim 5 \times 10^7  ~M_\odot$/sq deg. Considering the total region covered within 1.6 deg of the Galactic Centre (the nuclear bulge of the MW), if we assume that the RRab volume density is constant, we obtain a total number of $N \sim 13000$ RRab inside this area. Obviously, this is an unreasonable assumption, but we take the flat density profile as a lower limit. Using this number we obtain a total mass of $>6 \times 10^8 ~M_\odot$ for the old metal-poor stars inside 1.6 deg. A more reasonable assumption in order to make an extrapolation into the inner regions is that the RRab  keep their steep $r^{-3.5}$  density profile, as observed by the MACHO and OGLE microlensing experiments in the Galactic bulge \citep{minniti98,pietrukowicz15}. Using this steep density profile we obtain a total number of $N\sim 31,000$ RRab for the region with $R_G<1.6$ deg, from which we infer a total mass of $>1.4 \times 10^9 ~M_\odot$ for the old metal-poor stars. This is a large total mass that was obtained under two important assumptions: an $\omega$ Cen-like population, and an inner $r^{-3.5}$ density profile. These assumptions may be incorrect, and this number may be off by a factor of a few. However, we state that this total estimated mass would not be wrong by an order of magnitude in either direction. In fact, we can also argue that it is a lower limit because our sample is incomplete (at least 10\% incomplete due to the lack of low-amplitude RRab with $A(K_s)<0.15$ mag, as discussed above) even in the less reddened region at $1.6$ deg from the Galactic Centre that we used to zero-point our density law. 

Even though it is relatively small compared with the total mass of the Galactic bulge \citep[$M_{bulge}=2\times 10^{10} M_\odot$;][]{valenti16}, this mass of $>1.4 \times 10^9 ~M_\odot$ is a sizable contribution to the total mass budget of the Galactic nuclear bulge inside $R_G=1.6$ deg. This mass indicates that the old and metal-poor population may have played a dominant role in the inner dynamics of the MW, and also in the formation of the nuclear star cluster that contains the central black hole \citep[BH;][]{mastrobuonobattisti12}. In fact, the central BH weights a small fraction of the mass present today in old and metal-poor stars: $M_{BH}\simeq 4.2\times 10^6 ~M_\odot$ \citep{gillessen09,ghez08}.
One important implication would be that if the central BH formed very early, at the same time as the RRab studied here, and from similar material, its angular momentum may have been very low when it formed. If it has not spun up significantly by accreting disk meterial, a non-rotating supermassive BH would explain the absence of a strong jet, that is present in other supermassive BHs at the centres of distant galaxies.
\\

\section{Proper Motions}
\label{s_pms}
In this section we discuss the PM measurements for our RRab sample. These PMs have been measured following the procedures of \cite{contrerasramos17}, where a detailed description is given. The PMs computed in the present work are relative to the bulge population, obtained using red giant branch as astrometric reference stars. The PMs are more complicated to obtain than our previous works because of the high field density, however, we achieve a mean precision 
which suffices for our purposes of checking the validity of the sample and for adding an extra criterion for statistically discriminating foreground from distant RRab stars.

\begin{figure}[h]
\centering
\includegraphics[width=0.5\hsize]{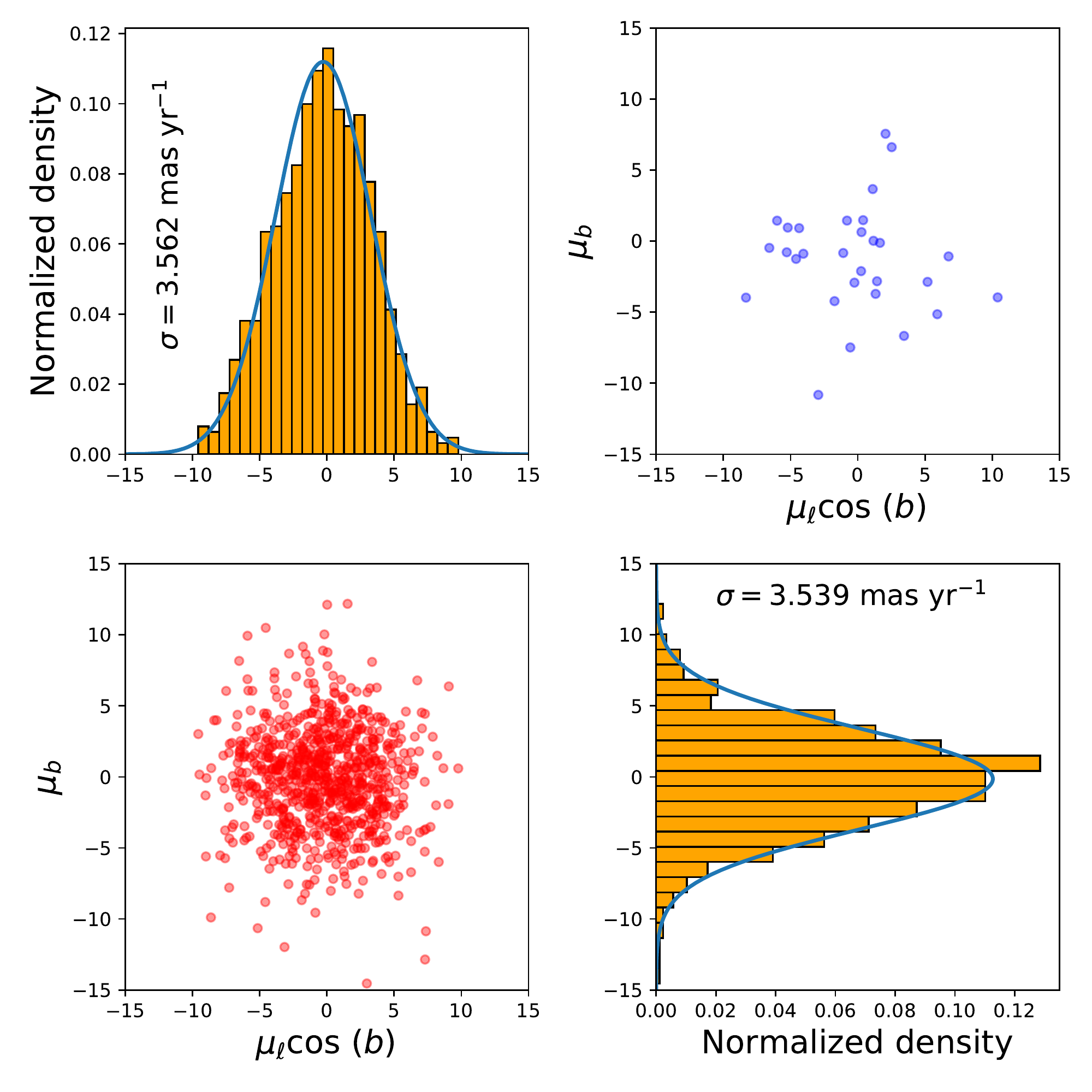}
\caption{Bottom Left panel: VPD in Galactic coordinates $\mu_{l}\cos(b)$ $vs$ $\mu_b$ in mas/yr for the candidate Galactic Centre RRab stars. Top Right panel: Similar VPD for the stars classified as foreground objects. Top Left: Galactic longitude PM distribution $\mu_{l}\cos(b)$ in mas/yr for the candidate Galactic Centre RRL. Bottom Right: Same as top left panel for the Galactic latitude PM distribution $\mu_b$ in mas/yr.}
\label{f_pm}
\end{figure}

Figure~\ref{f_pm} shows the PMs results in Galactic coordinates $\mu_{l}\cos(b)$ $vs$ $\mu_b$ in mas/yr for the candidate Galactic Centre RRab stars. The total number of RRab with relatively good PM measurements (total statistical error $< 3$ mas/yr) is 816, where we have also excluded 5 objects for which the PMs were impossible to measure because of crowding, and the 30 RRL deemed to be foreground/background objects (see bottom panel of Figure~\ref{f_KvsWK}). The lower left panel presents the vector point diagram (VPD) for the RRab sample showing a symmetric distribution, indicating a spherically symmetric spatial distribution. In particular, the Galactic longitude PM distribution does not have larger width that the Galactic latitude PM distribution, as one would expect from a flattened distribution or from a rotating population. This effect is even clearer in the lower right and top left panels of Figure~\ref{f_pm}, where the PM distribution in the Galactic longitude and latitude $\mu_{l}\cos(b)$ and $\mu_b$ in mas/yr, are shown. No significant asymmetries are observed in either of these distributions. We measure $\sigma_{\mu_l\cos(b)}=3.562 \pm 0.09$ mas/yr, and $\sigma_{\mu_b}=3.539 \pm 0.09$ for $N=816$ sample RRab stars with good PM measurements. The median statistical error in our PM measurements is 1.3 mas/yr, and considering 0.37 mas/yr due to systematic errors \citep[][]{contrerasramos17} yields a total error budget of 1.35 mas/yr in each PM coordinate.
After subtracting these errors in quadrature, we obtain $\sigma_{\mu_l\cos(b)}=3.29$ mas/yr, and $\sigma_{\mu_b}=3.27$ mas/yr, respectively. This allows us to conclude that the PM dispersions are real, that the RRab population in the Galactic Centre region has intrinsically a high velocity dispersion. In order to compute the tangential velocity dispersion we use $V_T$ (km/s) $= 4.74~\times$ PM (arcsec/yr) $\times~d$ (pc). 
At the distance of $D_{gc}=8.0$ kpc, these quantities correspond to tangential velocity dispersions of $\sigma V_l = 125.0 \pm 3.4$ km/s and $\sigma V_b = 124.1\pm 3.4$ km/s, respectively. Adopting a larger distance to the Galactic Centre, $D_{gc}=8.3$ kpc \citep{dekany13}, results in $\sigma V_l = 129.7 \pm 3.5$ km/s and $\sigma V_b = 128.7 \pm 3.5$ km/s, respectively. This very high tangential velocity dispersion in both components (along Galactic longitude and latitude) is not surprising, in light of the new results of \cite{zoccali17} and \cite{valenti16}, who found a central peak in the radial velocity dispersion in the inner bulge. In their innermost field at $1.0$ deg from the Galactic Centre they measure $\sigma RV\simeq 125$ and $145$ km/s for their metal-poor and metal-rich population of red giants, respectively. Note that they defined metal-poor as all the red giants with $-0.8<[Fe/H] <-0.1$ dex, and the metal-rich giants with $[Fe/H]>0.0$. Considering that the RRab measured here are the most extreme tail of the metal-poor distribution, it is then not surprising that they exhibit a large velocity dispersion. This is a useful comparison with the red giants, because unfortunately, the radial velocities would be difficult to measure for our sample RRab.



\begin{figure}[h]
\centering
\includegraphics[width=1.0\hsize]{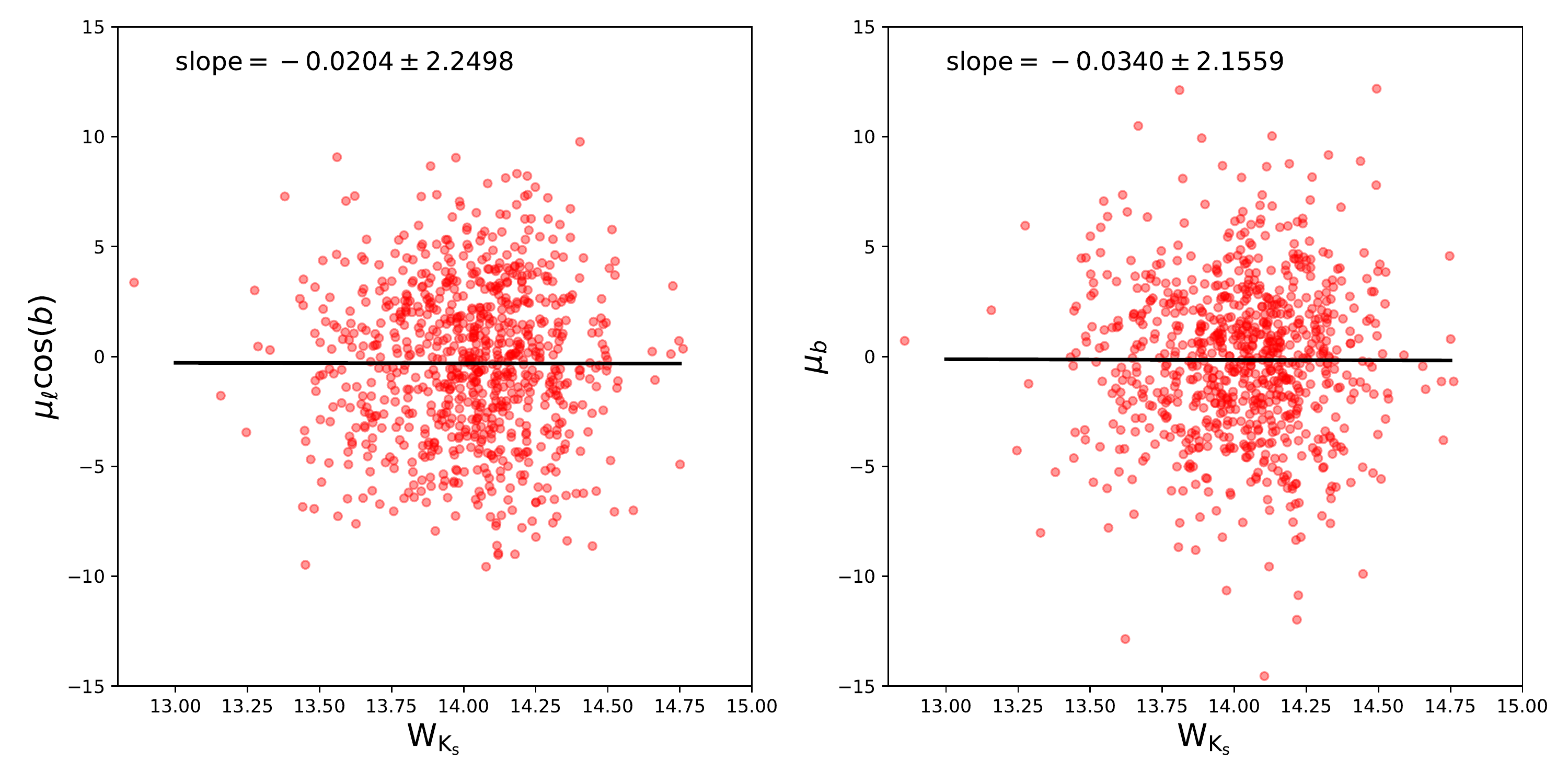}
\caption{\textbf{Left: $W_{K_s}$-band magnitude for the candidate Galactic Centre RRL as a function of Galactic longitude proper motion $\mu_{l}\cos(b)$ in mas/yr. Right: Same as left panel for $W_{K_s}$-band magnitude $vs$ the Galactic latitude proper motion $\mu_b$ in mas/yr. The slope of the distributions suggest that there is no dependence of PM with distance from de Galactic Centre.}}      
\label{f_pm_2}
\end{figure}

Figure~\ref{f_pm_2} plots the reddening-corrected Wesenheit magnitude dependence of the measured $\mu_{l}\cos(b)$ and $\mu_b$ PMs in mas/yr. This figure illustrates that there is not a strong dependence of PM with magnitude for the Galactic Centre RRab. The computed slope in the case of $W_{K_s}$ $vs$ $\mu_{l}\cos(b)$ also shows that there is no evidence for rotation, given that the brighter (closer) RRab have similar motions as the fainter (more distant) RRab stars. 

While the PMs are not accurate enough to explore in detail the kinematics around the Galactic Centre, they clearly show that there is no significant rotation of the population, where one would expect the PM in the Galactic longitude direction to be larger than the PM in the latitude direction. In fact, the PM distribution shows in Figure~\ref{f_pm} is round, and we find that $\sigma_{\mu_l\cos(b)} \simeq \sigma_{\mu_b}$, indicating a fairly isotropic velocity distribution for our RRL sample. Our results are in good agreement with the study of radial velocities of BRAVA RRL stars at lower latitudes \citep{kunder16}. 
 
We note that these RRL are fundamentally different in properties than the red giants in the same region, that have been found to be divided into two main populations: one metal rich component with $[Fe/H]\sim +0.3$ dex that follows a barred distribution, and another more  metal-poor component with mean $[Fe/H]\sim -0.4$ dex that has a more spherical distribution, both populations showing not negligible net rotation \citep{zoccali17}. The RRL here are more metal-poor in the mean and show zero rotation, suggesting to be more consistent with the extension of the old halo populations into the inner regions \cite{minniti96}.
This is our second major result: the RRL population in the Galactic Centre region shows very high tangential velocity dispersion of $\sigma V_l = 125.0$ km/s and $\sigma V_b = 124.1$ km/s and not bulk rotation.
The RRab are a very special population, because they represent the oldest and most metal-poor tracers that one can measure. The additional fact that the net angular momentum of the old and metal-poor stellar population is zero is also important. This has profound implications for the formation of the central regions of the MW, and the Galactic Centre in particular. This supports the dissipational collapse formation scenario of this region.
\\

\section{Discussion} 
\label{s_populations}
Are the RRL in the vicinity of the Galactic Centre noteworthy in any way? In this Section we discuss briefly the comparison with the RRL populations of other regions and components of the MW. We address the similarities and differences with the RRab found in the Galactic disk, halo, and inner and outer bulge, in order to discern if the Galactic Centre RRab come from a different population.

(1) the inner bulge RRab: The microlensing surveys have discovered thousands of RRL stars in the inner bulge. In particular, the MACHO and OGLE surveys show similar distance and metallicity properties as our sample RRab \citep{alcock98,kunder09,pietrukowicz15}. These samples, however, do not overlap because our RRL are located in very heavily reddened regions that were not accessible to the optical surveys.

(2) the outer bulge RRab: \cite{gran16} found a thousand RRab in the bulge-halo transition region, with projected distances between 1.2 and 1.4 kpc from the Galactic Centre. This sample also shows similar characteristics as our sample of Galactic Centre RRab. For example, their Bailey diagram looks qualitatively very similar. The Galactic Centre RRL however, have slightly shorter mean periods, suggesting a more metal-rich population that the outer bulge RRL.

(3) the globular cluster $\omega$\,Cen: \citet{navarrete15} and \citet{braga18} recently completed the NIR census and analysis of the RRL population associated with $\omega$\,Cen, the most massive Galactic globular cluster. $\omega$\,Cen is the prototypical example of an Oosterhoff type I population. 
As a whole, the RRab stars of our sample do not follow the same loci in the Bailey diagram as the RRab that belong to this massive cluster. This is relevant because globular clusters may have been part of the building blocks of the inner MW \citep{capuzzodolcetta93}, but these buiding blocks should be different from $\omega$Cen-like globular clusters. 
However, \cite{minniti17_4} have used VVV data in order to search for concentrations of RRab that would belong to unidentified bulge globular clusters. They were able to identify a dozen candidate globular clusters, which of course need to be confirmed with radial velocity measurements for example.

(4) the VVV disk sample: Recently \cite{minniti17_2} presented a large sample of RRab located in the plane of the MW. Collectively, these stars have are also similar properties (like mean periods, amplitudes and metallicities) to our RRab in the Galactic Centre.

(5) the halo population: There are various recent surveys of the RRab population in the Galactic halo \citep[e.g.][and references therein]{torrealba15}. The main differences that we find with the RRab in the Galactic Centre is that the outer halo RRab have a higher relative number of OoII RRab than our sample, and that the mean periods of the Galactic Centre RRab are shorter than the halo samples, indicating a more metal-rich population. The inner halo, however, seems to have more similar global properties as the Galactic bulge RRab \citep{abbas14}.
\\

\section{Conclusions}
\label{s_conclusion}
The first RRL in the Galactic Centre region have been discovered and initially characterized by \cite{minniti16} and \cite{dong17}. In this work we present an extended sample based on the VVV survey NIR photometry.
We have discovered 948 new RRab candidate variable stars within 100 arcmin of the Galactic Centre. We measured accurate positions, NIR mean magnitudes, colors, periods, amplitudes, and relative PMs, as well as estimate distances and rough metallicities for the whole sample. 

The colour-magnitude diagrams reveal the dramatic effects of high and variable extinction in the Galactic Centre region. Adopting an intrinsic mean colors for the RRab stars $J-K_{s}=0.17 \pm 0.03$ and judging from the measured NIR colors of the present RRL sample ($0.6<J-K_{s}<4.27$), the reddening in this field ranges from $E(J-K_{s})=0.44$ to $4.1$, and the extinction ranges from $A_{K_s}=0.19$ to $1.75$. We conclude that this is the major limitation for the discovery of RRL in this region. The reddening slope using the RRab sample is measured to be $A_{K_s}/E(J-K_{s})=0.438\pm 0.016$, which is consistent with recent measurements using VVV data \citep[0.428;][]{alonsogarcia17}.

We measure individual distances using the VVV NIR photometry. The distance distribution shows that the bulk of the sample is located at the distance of the bulge, as expected. We use the $P-W_{K_s}$ relation to separate 25 RRL located outside of the bulge, at a distance within 5 kpc from the Sun. We also identify five distant RRL, located behind the bulge, at $D\approx 12$ kpc, indicating that there are regions where the extinction is less severe. 
The mean distance for our RRab sample is $D=8.05 \pm 0.024$ kpc. 

The observed density of RRab variables in the Galactic Centre region is very high (1000 RRL/sq deg at $R=1.6$ deg). This agrees with the extrapolation of the RR Lyr density profile from the bulge microlensing experiments into the inner regions \citep{minniti98,pietrukowicz15}, and implies that the old metal-poor populations make a significant contribution to the total mass budget in this region.

The mean period is $P=0.5446 \pm 0.0025$ days, yielding  a mean metallicity of $[Fe/H] = -1.30 \pm 0.01$ ($\sigma=0.33$) dex for the RRab sample in the Galactic Centre region. This sample is compared with the VVV RRL samples from the MW disk \citep{minniti17_2}, middle bulge \citep{pietrukowicz15} and outer bulge \citep{gran16}. These comparisons show reasonable agreement, and we do not find major differences among these different samples. However, the period distribution suggests a higher metallicity than the Oosterhoff type I population in the outer halo studied by \cite{torrealba15}. The Bailey diagram shows also a predominant population that is more extreme than Oosterhoff type I RRL stars.

We measure PMs, firmly establishing the kinematics of the old and metal-poor population in the Galactic Centre region. We find a very large velocity dispersion $\sigma \approx 130$ km/s for the RRab variable stars, and no indication of significant rotation. The PM kinematics shows that this specific bulge population may have formed by dissipational collapse, and the kinematics are also consistent with the scenario that the population may be an extension of the inner halo \citep{minniti96}. The results are also consistent with the observed chemical history, as the bulge red giants seem to exhibit a rapid  formation judging from their measured chemical enrichment  \citep[e.g. O-enhancement found by][and references therein]{zoccali08}. The measurement of radial velocities for the present sample would be very profitable to investigate the orbital dynamics of this old and metal poor population. These RVs would also be useful to confirm or disproof a number of globular cluster candidates recently identified in the Galactic Centre region as concentrations of RRab \citep{minniti17_4}.

The RRL population of the Galactic Centre deserves more observations and theoretical modelling to see in what proportion it could come from disrupted globular clusters, what fraction of the population is inherent to the bulge, and what fraction is just passing by from the halo population. Indeed, these are useful prime targets for future spectroscopy, in order to investigate their 3-D kinematics and chemical compositions. In addition, the present work underscores the need for the WFIRST space mission, that will be able to complete the RRL census that cannot be done from the ground in the most obscured regions. Without WFIRST this part of the Galaxy will remain hidden.
\\

\acknowledgments
We gratefully acknowledge the use of data from the ESO Public Survey program ID 179.B-2002 taken with the VISTA telescope, and data products from the Cambridge Astronomical Survey Unit (CASU).  Support for the authors is provided by the BASAL Center for Astrophysics and Associated Technologies (CATA) through grant PFB-06, and the Ministry for the Economy, Development, and Tourism, Programa Iniciativa Cientifica Milenio through grant IC120009, awarded to the Millennium Institute of Astrophysics (MAS). D.M., and M.Z. acknowledge support from FONDECYT Regular grants No. 1170121, and 1150345, respectively. P.H. acknowledges financial support from FONDECYT regular grant 1170305. F. G. acknowledge support from CONICYT-PCHA Doctorado Nacional 2017-21171485 and Proyecto Fondecyt Regular 1150345. J.A-G. acknowledges support by FONDECYT Iniciación 11150916. DM is also grateful for the hospitality of the Vatican Observatory. \\

\appendix
\label{appendixA}
\section{Positive-Unlabeled Random Forest classifier}
Positive-Unlabeled (PU) learning refers to the case where a set representing a known class (P) and a usually larger set containing unlabeled data (U) are used to train a semi-supervised classifier to retrieve the objects in U that are most similar to P. In this case P and U correspond to the RR Lyrae found in \cite{gran16} and the light curves of the bulge, respectively. We draw from \cite{mordelet14} and train an ensemble of decision trees using a PU version of the bagging algorithm (a python implementation of the algorithm is at \url{www.github.com/phuijse/bagging_pu}). To obtain features the light curves are (1) folded, (2) aligned in phase, (3) normalized and (4) interpolated to a 50-points regular grid using Gaussian Process (GP) regression. For U the period is found using mutual information \citep{huijse17}. Fifty additional features folding with twice the reported period are also included. The 100 GP features plus the amplitude and period are the input to the classifier. This differs from \cite{elorrieta16} in that the classifier uses the distribution of U (semi-supervision) and in the features that characterize the light curves.

\newpage

\bibliography{biblio}

\floattable
\begin{deluxetable}{lrcccccccccc}
\tablecaption{RR Lyrae sample photometric observations and relative proper motions
\label{t_tab1}}
\tablecolumns{12}
\tablenum{1}
\tablewidth{0pt}
\tablehead{
\colhead{ID} & \colhead{L} & \colhead{B} & \colhead{$K_{s}$} & \colhead{$H$} & \colhead{$J$} & \colhead{$Y$} & \colhead{$Z$} & \colhead{$\mu_{l}\cos(b)$} & \colhead{$\epsilon\mu_{l}\cos(b)$} & \colhead{$\mu_{b}$} & \colhead{$\epsilon\mu_{b}$} \\
\colhead{} & \colhead{(deg)} & \colhead{(deg)} & \colhead{} & \colhead{} & \colhead{} & \colhead{} & \colhead{} & \colhead{(mas/yr)} & \colhead{} & \colhead{(mas/yr)} & \colhead{}
}
\startdata
b305\_209\_26117 & 359.5813 & -1.5817 & 14.636 & 14.895 & 15.527 & 16.342 & 16.983 & 3.713 & 1.04 & -2.362 & 1.16 \\
b305\_209\_26795 & 359.5373 & -1.5830 & 14.534 & 14.731 & 15.429 & 16.072 & 16.688 & -2.506 & 1.30 & 0.182 & 0.86 \\
b305\_209\_43094 & 359.5522 & -1.6156 & 14.528 & 14.742 & 15.429 & 16.254 & 16.917 & -3.284 & 1.26 & -0.652 & 1.64 \\
b305\_209\_58529 & 359.6060 & -1.6467 & 14.263 & 14.475 & 15.044 & 15.758 & 16.344 & -0.102 & 1.03 & -0.586 & 0.75 \\
b305\_209\_66679 & 359.6713 & -1.6635 & 14.305 & 14.577 & 15.253 & 15.690 & 16.246 & -1.497 & 1.04 & -0.206 & 1.09 \\
b305\_213\_10296 & 359.9688 & -1.5532 & 14.279 & 14.589 & 15.242 & 15.970 & 16.586 & -1.395 & 1.35 & 3.026 & 1.44 \\
b305\_213\_21036 & 359.9372 & -1.5749 & 14.785 & 15.060 & 15.581 & 16.539 & 17.095 & -0.618 & 1.77 & -5.732 & 1.67 \\
b305\_213\_34987 & 0.0358 & -1.6043 & 14.563 & 15.101 & 15.733 & 16.163 & 16.767 & -0.115 & 1.42 & 0.767 & 1.47 \\
b305\_213\_60049 & 359.8865 & -1.6542 & 14.608 & 14.811 & 15.451 & 15.880 & 16.407 & 3.444 & 1.83 & -5.904 & 1.34 \\
b305\_213\_60598 & 359.9802 & -1.6558 & 14.001 & 14.399 & 14.943 & 15.331 & 15.908 & -7.267 & 0.71 & -7.795 & 1.25 \\
b305\_213\_68937 & 0.0212 & -1.6727 & 14.480 & 14.668 & 15.285 & 15.982 & 16.511 & -3.544 & 1.25 & 3.481 & 1.14 \\
b305\_213\_81108 & 359.9462 & -1.6965 & 14.346 & 14.552 & 15.096 & 15.783 & 16.297 & -1.017 & 1.25 & -1.107 & 1.01 \\
\enddata
\tablecomments{Table \ref{t_tab1} is published in its entirety in the machine readable format.  A portion is shown here for guidance regarding its form and content.}
\end{deluxetable}
\floattable
\begin{deluxetable}{lccccccc}
\tablecaption{Measured stellar parameters
\label{t_tab2}}
\tablecolumns{6}
\tablenum{2}
\tablewidth{0pt}
\tablehead{
\colhead{ID} & \colhead{$E(J-K_s)$} & \colhead{$A_{K_s}$} & \colhead{$(m-M)_0$} & \colhead{$D_{\odot}$} & \colhead{$P$} & \colhead{$A(K_s)$} & \colhead{$[Fe/H]$}\\ 
\colhead{} & \colhead{} & \colhead{} & \colhead{} & \colhead{(kpc)} & \colhead{(days)} & \colhead{} & \colhead{(dex)}
}
\startdata
b305\_209\_26117 & 0.721 & 0.309 & 14.698 & 8.703 & 0.54339 & 0.298 & -1.32 \\ 
b305\_209\_26795 & 0.725 & 0.310 & 14.856 & 9.359 & 0.68971 & 0.260 & -1.90 \\ 
b305\_209\_43094 & 0.731 & 0.313 & 14.473 & 7.846 & 0.49046 & 0.304 & -1.07 \\ 
b305\_209\_58529 & 0.611 & 0.262 & 14.547 & 8.118 & 0.63748 & 0.224 & -1.71 \\ 
b305\_209\_66679 & 0.778 & 0.333 & 14.320 & 7.312 & 0.53216 & 0.308 & -1.27 \\ 
b305\_213\_10296 & 0.793 & 0.339 & 14.459 & 7.795 & 0.62181 & 0.228 & -1.65 \\ 
b305\_213\_21036 & 0.626 & 0.268 & 14.820 & 9.203 & 0.51083 & 0.306 & -1.17 \\ 
b305\_213\_34987 & 1.000 & 0.428 & 14.650 & 8.513 & 0.62012 & 0.277 & -1.64 \\ 
b305\_213\_60049 & 0.673 & 0.288 & 14.541 & 8.093 & 0.47411 & 0.294 & -0.99 \\ 
b305\_213\_60598 & 0.772 & 0.331 & 14.066 & 6.503 & 0.55565 & 0.254 & -1.38 \\ 
b305\_213\_68937 & 0.635 & 0.272 & 14.507 & 7.970 & 0.50876 & 0.285 & -1.16 \\ 
b305\_213\_81108 & 0.580 & 0.248 & 14.481 & 7.875 & 0.54951 & 0.264 & -1.35 \\ 
\enddata
\tablecomments{Table \ref{t_tab2} is published in its entirety in the machine readable format.  A portion is shown here for guidance regarding its form and content.}
\end{deluxetable}

\end{document}